\documentclass{aa}

\usepackage{graphicx,natbib}
\usepackage{bm}

\usepackage{color}

\begin{document}

   \title{Pulse-beam heating of deep atmospheric layers triggering their oscillations and upwards moving shocks that can modulate the reconnection in solar flares}

   \authorrunning{P. Jel\'\i nek et al.}
   \titlerunning{Pulse-beam heating of deep atmospheric layers triggering their oscillations ...}

   \author{P. Jel\'\i nek\inst{1},
           M. Karlick\'y\inst{2} 
           }
   \offprints{P. Jel\'\i nek, \email{pjelinek@prf.jcu.cz}}

   \institute{University of South Bohemia, Faculty of Science, Institute of Physics, Brani\v sovsk\'a 1760, CZ -- 370 05 \v{C}esk\'e
             Bud\v{e}jovice, Czech Republic
             \and
             Astronomical Institute of the Czech Academy of Sciences, Fri\v{c}ova 258, CZ -- 251 65 Ond\v rejov, Czech
             Republic\\
    }
   \date{Received ; accepted }


  \abstract
   {}
   {We study processes occurring after a sudden heating of the chromosphere at the flare arcade footpoints
    which is assumed to be caused by particle beams.}
   {For the numerical simulations we adopt a 2-D magnetohydrodynamic (MHD) model, in which we solve a full set of the
   time-dependent MHD equations by means of the FLASH
   code, using the Adaptive Mesh Refinement (AMR) method.}
   {In the initial state we consider a model of the solar atmosphere with densities according
   to the VAL-C model and the magnetic field arcade having the X-point structure above, where the magnetic reconnection is assumed.
   We found that the sudden pulse-beam heating of the chromosphere at the flare arcade footpoints generates magnetohydrodynamic shocks,
   one propagating upwards and the second one propagating
   downwards in the solar atmosphere. The downward moving shock is reflected at deep and dense atmospheric
   layers and triggers oscillations of these layers. These oscillations generate
   the upwards moving magnetohydrodynamic waves that can influence the above located magnetic reconnection in a quasi-periodic way.
   Because these processes require a sudden heating in very localized regions in the chromosphere therefore they can
   be also associated with seismic waves.}
   {}

   \keywords{Sun: flares -- Sun: oscillations -- magnetohydrodynamics (MHD) -- methods: numerical}

   \maketitle

\section{Introduction}

In solar flares oscillations are quite typical. They are observed in a broad
range of the electromagnetic emission from radio, soft X-ray, hard X-ray,
ultraviolet up to gamma-rays, see
e.g.~\citep{1984ApJ...279..857R,2003SoPh..218..183F,2005A&A...435..753W,2006A&A...452..343N,2010PPCF...52l4009N}.
These oscillations are with periods from sub-seconds to tens of
minutes~\citep{2006A&A...460..865M,2008SoPh..253..117T,2010SoPh..261..281K,2010SoPh..267..329K,2014ApJ...791...44H,2014A&A...569A..12N}.
Many models of these observations have been proposed, see e.g. our last papers
\citep{2013MNRAS.434.2347J,2016SoPh..291..877G,2017ApJ...847...98J} and for the review see e.g.
the papers by
\citep{2014RAA....14..805P,2016SSRv..200...75N,2018SSRv..214...45M}.

It is commonly accepted that during the impulsive phase of solar flares,
particle beams, which are accelerated by the magnetic reconnection in the low
corona, propagate downwards along the legs of flare arcade and bombard dense
chromospheric layers at their footpoints. Due to this bombardment the
chromosphere at arcade footpoints is rapidly heated and the hard X-ray emission
and magnetohydrodynamic shocks are
generated~\citep{1971SoPh...18..489B,1984SoPh...90..357M,1985SoPh...96..317M,1985ApJ...289..434F,1985ApJ...289..425F,1985ApJ...289..414F,1989ApJ...341.1067M,1990SoPh..130..347K,
1992A&A...264..679K,1994ApJ...426..387H,1999ApJ...521..906A,2005ApJ...630..573A,2014A&A...563A..51V,2016A&A...590A...4K}.

In this paper we present new aspects of the above mentioned processes. Namely
as will be shown in the following, a sudden and very localized heating of the
chromosphere at the flare arcade footpoints triggers not only upwards and
downwards moving shocks, but also the oscillations of deep atmospheric layers
that then quasi-periodically generate the upwards moving shocks/waves which can
modulate the primary-flare reconnection process located above. The reconnection
rate is then quasi-periodically modulated and thus the oscillations found in
the flare emissions can be explained. This scenario is a new alternative of the
model presented by ~\cite{2006A&A...452..343N}, where the authors assumed a
huge oscillating loop (resonator) generating waves which modulate a nearby
flare magnetic reconnection.

The structure of the present paper is as follows. In Section 2 we present our numerical
model, including the governing equations, initial equilibrium and perturbations. The results of
numerical simulations and their interpretation are summarized in Section 3.
Finally, we complete the paper by conclusions in the last Sect. 4.

\section{Model}
\subsection{Governing equations}
In our computer simulation we implement a gravitationally stratified solar
atmosphere, according to VAL-C model, in which the plasma dynamics are
described by the two-dimensional (2-D), time-dependent non-ideal (resistive)
magnetohydrodynamic (MHD) equations. We solve the problem numerically with the
use of FLASH code~\citep{2013JCoPh.243..269L}, where the MHD equations are
formulated in conservative form as follows
\begin{equation}\label{eq1}
\frac{\partial \varrho}{\partial t} + \nabla \cdot(\varrho \bm{\mathrm{v}}) = 0,
\end{equation}
\begin{equation}\label{eq2}
\frac{\partial \varrho \bm{\mathrm{v}}}{\partial t} + \nabla \cdot \left(\varrho \bm{\mathrm{v}} \bm{\mathrm{v}} -\bm{\mathrm{B}}\bm{\mathrm{B}}\right) + \nabla p_{*}=\varrho \bm{\mathrm{g}},
\end{equation}
\begin{eqnarray}\label{eq3}
\frac{\partial \varrho E}{\partial t} + \nabla \cdot \left[\left(\varrho E+p_{*}\right)\bm{\mathrm{v}} - \bm{\mathrm{B}}(\bm{\mathrm{v}} \cdot \bm{\mathrm{B}})\right] =  \nonumber \\
= \varrho \bm{\mathrm{g}} \cdot \bm{\mathrm{v}} + \nabla \cdot (\bm{\mathrm{B}} \times (\eta \nabla \times \bm{\mathrm{B}})),
\end{eqnarray}
\begin{equation}\label{eq4}
\frac{\partial \bm{\mathrm{B}}}{\partial t} + \nabla \cdot (\bm{\mathrm{v}}\bm{\mathrm{B}}-\bm{\mathrm{B}}\bm{\mathrm{v}})=-\nabla \times (\eta \nabla \times \bm{\mathrm{B}}),
\end{equation}
\begin{equation}\label{eq5}
\nabla\cdot\bm{\mathrm{B}}=0.
\end{equation}
Here $\varrho$ is the mass density, $\bm{\mathrm{v}}$ is the flow velocity,
$\bm{\mathrm{B}}$ is the magnetic field strength, $\bm{\mathrm{g}} =
[0,-g_{\sun},0]$ is the gravitational acceleration with $g_{\sun} =
274~\mathrm{m s^{-2}}$ and $\eta$ is the magnetic diffusivity
taken constant throughout of the numerical box and corresponding to the
relation $\eta = 10^{9}T^{-3/2}$ \citep{2014masu.book.....P} for the coronal
temperature $T = 10^6~\mathrm{K}$.

The total pressure $p_{*}$ is given by:
\begin{equation}\label{eq6}
p_{*} = \left(p + \frac{B^2}{2 \mu_0}\right),
\end{equation}
$p$ is the fluid thermal pressure, $B$ is the magnitude of the magnetic field. The specific total energy $E$ in Eq. (\ref{eq3}) is expressed as:
\begin{equation}\label{eq7}
E = \epsilon + \frac{v^2}{2} + \frac{B^2}{2 \mu_0 \varrho},
\end{equation}
where $\epsilon$ is the specific internal energy:
\begin{equation}\label{eq8}
\epsilon = \frac{p}{(\gamma-1)\varrho},
\end{equation}
with the adiabatic coefficient $\gamma = 5/3$, $v$ is the magnitude of the flow
velocity and $\mu_0 = 1.26 \times 10^{-6}~\mathrm{H m}^{-1}$ is the magnetic
permeability of free space.

\subsection{Initial state}
For a still ($\bm{\mathrm{v}} = \bm{0}$) equilibrium, the Lorentz and gravity forces have to be balanced
by the pressure gradient in the entire physical domain,
\begin{equation}\label{eq9}
-\nabla p+\bm{\mathrm{j}}\times\bm{\mathrm{B}} + \varrho \bm{\mathrm{g}} = \bm{\mathrm{0}}.
\end{equation}
Assuming a force-free magnetic field, $\bm{\mathrm{j}} \times \bm{\mathrm{B}} = \bm{\mathrm{0}}$, in the null-point the
solution of the remaining hydrostatic equation yields
\begin{equation}\label{eq10}
p_{\rm h}(y) = p_0 \exp\left[-\int\limits_{y_0}^{y} \frac{1}{\Lambda(\tilde{y})}\mathrm{d}\tilde{y}\right],
\end{equation}
\begin{equation}\label{eq11}
\varrho(y) = \frac{p(y)}{g_{\sun}\Lambda(y)}.
\end{equation}
Here
\begin{equation}\label{eq12}
\Lambda(y) = \frac{k_\mathrm{B}T(y)}{\overline{m}g_{\sun}}
\end{equation}
is the pressure scale-height which in the case of isothermal atmosphere
represents the vertical distance over which the gas pressure decreases by a
factor of $e\approx 2.7$, $k_\mathrm{B} = 1.38 \times 10^{-23}~\mathrm{J\cdot
K^{-1}}$ is the Boltzmann constant and $\overline{m} = 0.6\, m_\mathrm{p}$ is
the mean particle mass ($m_\mathrm{p} = 1.672 \times 10^{-27}~\mathrm{kg}$ is
the proton mass), $p_0 \approx 10^{-2}~\mathrm{Pa}$ in Eq. (\ref{eq9}) denotes
the gas pressure at the reference level $y_0$. In our calculations we set and
hold fixed $y_0 = 10~\mathrm{Mm}$. For the solar atmosphere we used the
temperature profile derived by \citep{2008ApJS..175..229A}. At the top of the
photosphere, which corresponds to the height of $y=0.5~\mathrm{Mm}$, the
temperature is $T(y)=5700~\mathrm{K}$. At higher altitudes, the temperature
falls down to its minimal value $T(y)=4350~\mathrm{K}$ at $y \approx
0.95~\mathrm{Mm}$. Higher up the temperature rises slowly to the height of
about $y=2.7~\mathrm{Mm}$, where the transition region is located. Here the
temperature increases abruptly to the value $T(y)=1.5~\mathrm{MK}$, at the
altitude $y=10~\mathrm{Mm}$, which is typical for the solar corona.

The solenoidal condition, $\nabla\cdot\bm{\mathrm{B}}=0$, is identically
satisfied with the use of the magnetic flux function, $\bm{\mathrm{A}}$, such
as
\begin{equation}\label{eq13}
\bm{\mathrm{B}} = \nabla \times \bm{\mathrm{A}}.
\end{equation}
Specifically, to represent the non-potential null-point we use
$\bm{\mathrm{A}} = [0,0,A_z]$, such as \citep{1997GApFD..84..245P,2015ApJ...812..105J}
\begin{equation}\label{eq14}
A_z = \frac{1}{4} B_{\mathrm{0}} [(\mathcal{I}_t - \mathcal{I}_z) y^2 - (\mathcal{I}_t + \mathcal{I}_z) x^2],
\end{equation}
which gives us for magnetic field components
\begin{equation}\label{eq15}
B_x(x,y) = \frac{B_0}{2}(\mathcal{I}_t - \mathcal{I}_z) y
\end{equation}
and
\begin{equation}\label{eq16}
B_y(x,y) = \frac{B_0}{2}(\mathcal{I}_t + \mathcal{I}_z) x.
\end{equation}
Here $\mathcal{I}_t$ is the threshold current which only depends on the
parameters associated with the potential part of the field and it is assumed to
be a constant in our calculations. The parameter $\mathcal{I}_z$ is the
magnitude of the current perpendicular to the plane of the null-point. Both
$\mathcal{I}_t$ and $\mathcal{I}_z$ are free parameters which govern the
magnetic field configuration, see \cite{1996PhPl....3..759P} for more details.
Magnetic field at the reference level is set and hold fixed as
$B_0=10~\mathrm{G}$.

The equilibrium gas pressure and mass density are computed according to the following equations, see \cite{2010ARep...54...86S}:
\begin{equation}\label{eq17}
p(x,y) = p_{\rm h} - \frac{1}{\mu_0}\left[\int\limits_{-\infty}^{x}\frac{\partial^2 A}{\partial y^2}\frac{\partial A}{\partial x}\mathrm{d}x +
\frac{1}{2}\left(\frac{\partial A}{\partial x}\right)^2\right],
\end{equation}

\begin{eqnarray}\label{eq18}
\varrho(x,y) = \varrho_{\rm h}(y) &+& \frac{1}{\mu_0 g_{\sun}}\Bigg\{\frac{\partial}{\partial{y}}\Bigg[\int\limits_{-\infty}^{x} \frac{\partial^2 A}{\partial y^2}\frac{\partial A}{\partial x}\mathrm{d}x + \nonumber \\
&+& \frac{1}{2}\Bigg(\frac{\partial A}{\partial x}\Bigg)^2\Bigg] - \frac{\partial A}{\partial y} \nabla^2 A\Bigg\}.
\end{eqnarray}
With the use of Eq.~(\ref{eq14}) in these general formulas we obtain the
expressions for the equilibrium gas pressure \citep{2015ApJ...812..105J}
\begin{equation}\label{eq19}
p(x,y) = p_h(y) - \frac{B_0^2}{4\mu_0} \mathcal{I}_z (\mathcal{I}_t + \mathcal{I}_z) x^2
\end{equation}
and mass density
\begin{equation}\label{eq20}
\varrho(x,y) = \varrho_h(y) + \frac{B_0^2}{2\mu_0 g} \mathcal{I}_z(\mathcal{I}_t - \mathcal{I}_z) y.
\end{equation}
The initial state, showing the distribution of gas pressure in logarithmic
scale and governing magnetic field (black lines) is illustrated in Fig. 1. The
red line represents the separatrix.
\begin{figure}[htb!]
\centering
\includegraphics[width=3.5in]{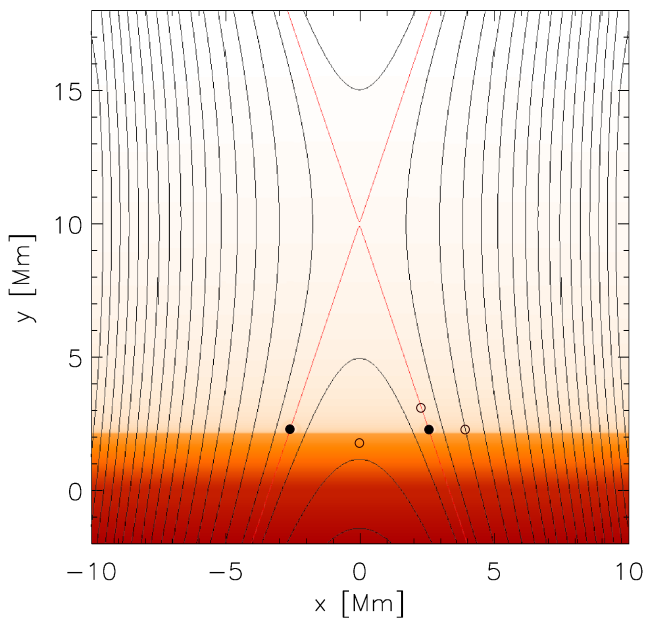}
\caption{Sketch of the initial state in the equilibrium represented by the distribution of gas pressure (in logarithmic scale)
and magnetic field lines (black lines) along with the separatrices (red lines).
The initial pressure pulses are depicted as two black full circles at separatrices
in the height $2.25~\mathrm{Mm}$ above the photosphere, i.e., at the transition region.
Three unfilled black circles represent the detection points.}
\label{fig1}
\end{figure}

Generally, the terms expressing the radiative losses
$R_{\mathrm{loss}}$, thermal conduction $T_{\mathrm{cond}}$ and heating $H$
should be added to the set of MHD equations. These terms are certainly
important in building a realistic numerical model. However, at this stage when
the wave and oscillatory processes are of primary interest, we neglect these
terms.

\subsection{Perturbations}

To generate the initial perturbation pulse, at the start of the numerical
simulation ($t = 0~\mathrm{s}$), the equilibrium in the transition region (TR)
is perturbed by the Gaussian pulse in the temperature (pressure) and has the
following form
\begin{eqnarray}\label{eq21}
p = p_0 \Bigg\{1 + A_\mathrm{p} \sum_{i=1}^2 \exp{\Bigg[-\frac{(x-x_i^{\mathrm{P}})^2+(y-y^{\mathrm{P}})^2}{\lambda^2} \Bigg]} \Bigg\},
\end{eqnarray}
where $p_0$ is the initial gas pressure, $A_\mathrm{p}$ is the initial
amplitude of the pressure pulse, $\lambda$ is the size of the pressure pulse,
$x_i^{\mathrm{P}}$ and $y^{\mathrm{P}}$ are the positions of the initial
perturbation pulses. The perturbation points are placed in $x_1^{\mathrm{P}} =
-2.58~\mathrm{Mm}$ and $x_2^{\mathrm{P}} = +2.58~\mathrm{Mm}$, respectively,
with $y^{\mathrm{P}} = 2.25~\mathrm{Mm}$ for both points, which corresponds to
the position of TR.

\subsection{Numerical code}
For solution of the MHD equations (\ref{eq1})-(\ref{eq4}) we use the FLASH
code, which is well tested, fully modular, parallel, multiphysics, open
science, simulation code that implements second- and third-order unsplit
Godunov solvers with various slope limiters and Riemann solvers as well as
adaptive mesh refinement (AMR)~\citep{2002cfd..book.....C}. The Godunov solver
combines the corner transport upwind method for multi-dimensional integration
and the constrained transport algorithm for preserving the divergence-free
constraint on the magnetic field~\citep{2009JCoPh.228..952L}. We have used the
minmod slope limiter and the Riemann solver e.g.,~\citep{2006IJNMF..52..433T}.
The main advantage of using AMR technique is to refine a numerical grid at
steep spatial profiles while keeping a grid coarse at the places where fine
spatial resolution is not essential. In our case, the AMR strategy is based on
controlling the numerical errors in a gradient of mass density that leads to
reduction of the numerical diffusion within the entire simulation region.

For our numerical simulations, we use a 2-D Eulerian box of its width $W =
20~\mathrm{Mm}$ and height $H = 20~\mathrm{Mm}$ as we set the numerical box as
$(-10,10)~\mathrm{Mm} \times (-2,18)~\mathrm{Mm}$ in $x$ and $y$ direction,
respectively. The spatial resolution of the numerical grid is determined with
the AMR method. We use the AMR grid with the minimum (maximum) level of the
refinement blocks set to 3 (7). The whole simulation region is covered by
$2962$ blocks. Since every block consists $8 \times 8$ numerical cells, this
number of blocks corresponds to $189568$ numerical cells and the smallest
spatial resolution is $\Delta x = \Delta y = 3.9~\mathrm{km}$.

At all boundaries, we fix all plasma quantities to their equilibrium values
using fixed-in-time boundary conditions, which lead only to negligibly small
numerical reflections of incident wave signals.

\section{Numerical results}

Prior to run of the numerical simulation, we verified that the system is in
equilibrium for the adopted grid resolution by running the system without any
pressure pulse. After this control test, we launched in the system two
symmetric pressure pulses with the amplitudes of $A_\mathrm{p} = 10.0$ in the
points at $x_{1,2}^{\mathrm{P}}$ and $y^{\mathrm{P}}$, see full black circles
in Fig. {\ref{fig1}}. By this pressure pulse, the temperature at these points
increased from the initial equilibrium value of $T_\mathrm{eq} \approx
0.17~\mathrm{MK}$ to the temperature $T \approx
1.7~\mathrm{MK}$.

\begin{figure*}[t!]
\centering
\includegraphics[width=7.25in]{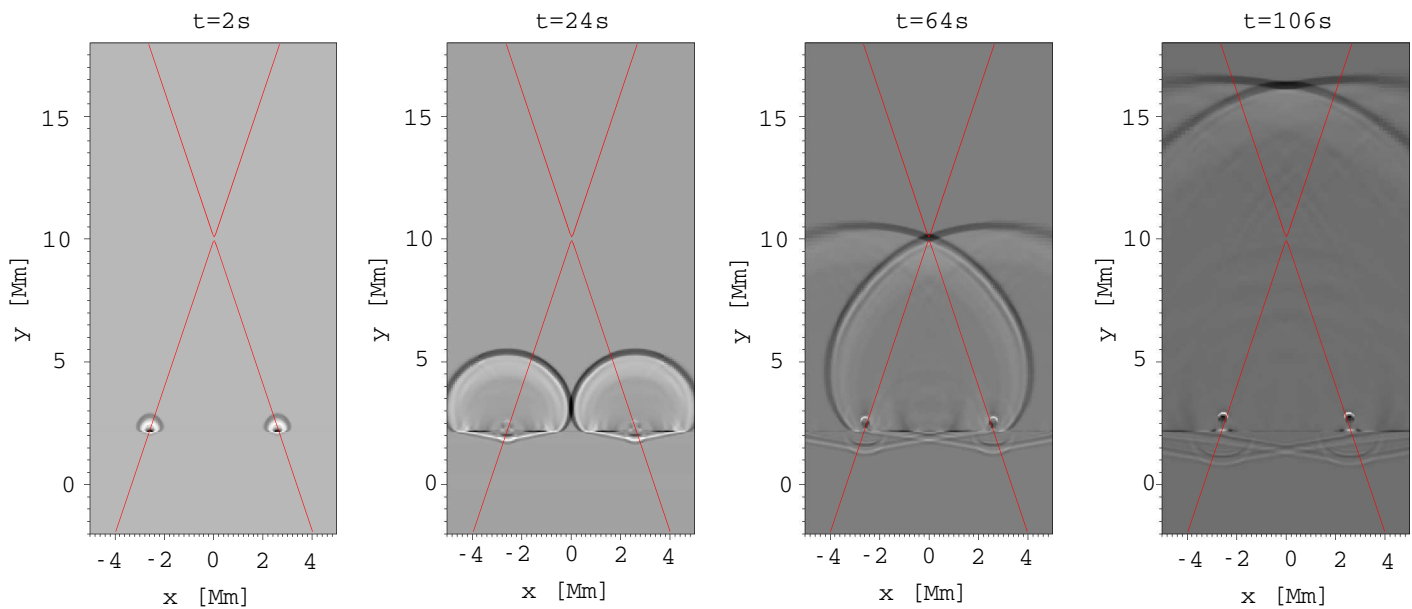}
\caption{Temporal evolution of changes of the mass density
$\Delta\varrho/\varrho_0$, at four different times, $t = 2, 26, 64$ and
$106~\mathrm{s}$. The initial pressure pulse was launched in TR on separatrices
(solid red lines).} \label{fig2}
\end{figure*}

In Figures {\ref{fig2}} -- {\ref{fig9}} we show the results from our numerical
simulations. For better readability and to see more details, Figs. {\ref{fig2}}
-- {\ref{fig4}} are shown within a range $(-5,5)~\mathrm{Mm}$ along
$x-$direction. 

\begin{figure*}[t!]
\centering
\includegraphics[width=7.25in]{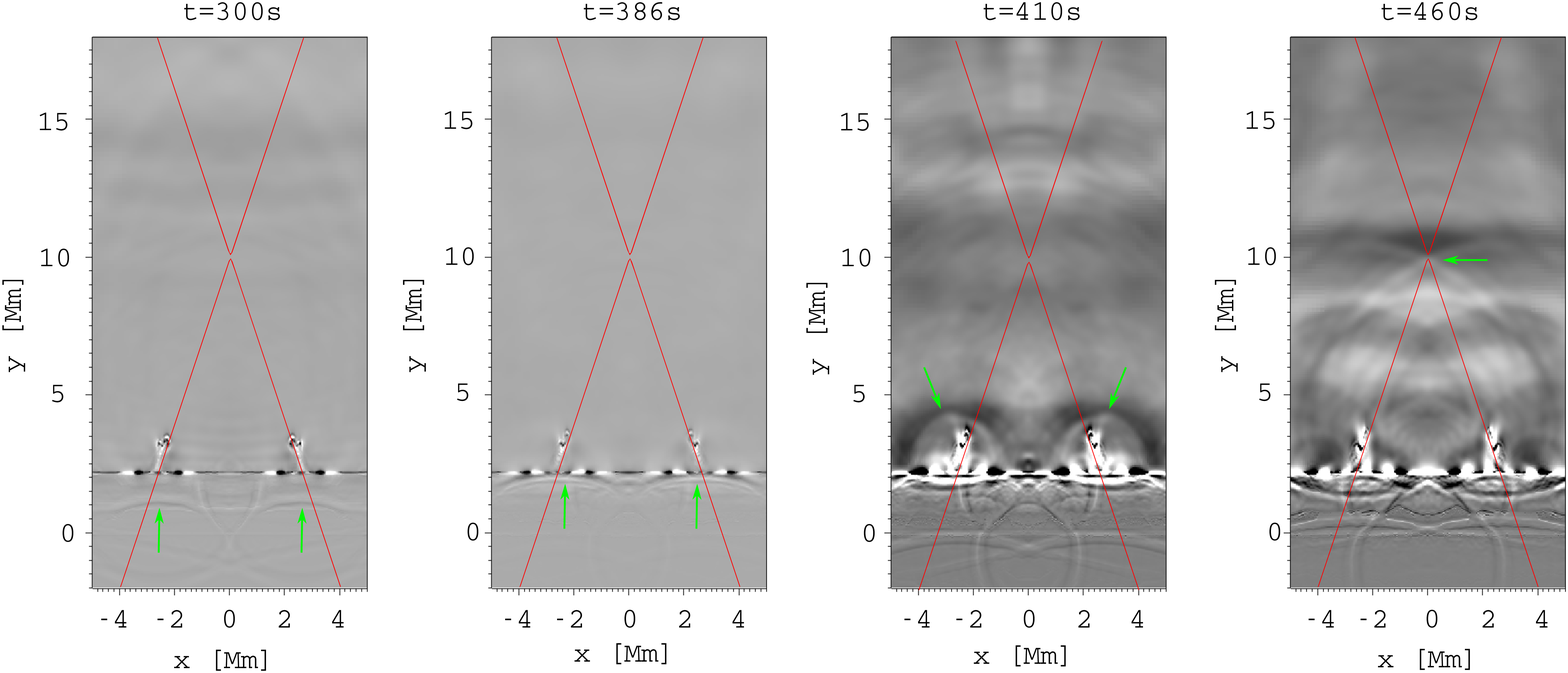}
\caption{Temporal ime evolution of changes of the mass density
$\Delta\varrho/\varrho_0$, at four different times, $t = 300, 386, 410$ and
$460~\mathrm{s}$, showing the secondary wave generated by the oscillation
of the photosphere The wave is marked by the lightgreen arrows, the red line
represents the separatrices.} \label{fig3}
\end{figure*}
In Figure {\ref{fig2}} we show the time evolution of changes of
the mass density $\Delta\varrho/\varrho_0$ for four different times, $t = 2,
26, 64$ and $106~\mathrm{s}$, respectively. The crossing of
the red lines indicates the X-point of the magnetic reconnection. At very
early phase of this time evolution we can see that the shock wave, generated by
the initial pressure pulse, propagates from both perturbation points
symmetrically and at time $t=26~\mathrm{s}$ both wavefronts come across each
other. Then both waves propagate towards the X-point ($x=0~\mathrm{Mm},
y=10~\mathrm{Mm}$) which they reach at  time $t=64~\mathrm{s}$. At time
$t\approx 106~\mathrm{s}$ the waves come to the upper boundary of the presented
region. It can also be seen that the waves propagate down to deeper layers of
the solar atmosphere. These waves are partially remain in the
photosphere and deeper layers of solar body and they are also partially
reflected with lower energy back, towards the higher altitudes in the solar
atmosphere, which is clearly seen in Fig. {\ref{fig3}}.

In Figure {\ref{fig3}} we present the time evolution of changes of the
mass density at later times, mainly at $t=300, 386, 410~\mathrm{s}$ and
$t=460~\mathrm{s}$. We can see here the wave reflected from the
photosphere ($y=0~\mathrm{Mm}$) as well as the waves propagating down to even
deeper layers and reflected later; see also last two right panels of Fig.
{\ref{fig2}} for times $t=64~\mathrm{s}$ and $t=106~\mathrm{s}$.  The reflected
waves are marked by lightgreen arrows. The figure at $t=300~\mathrm{s}$ shows
the wave shortly after the reflection from the photosphere. At time
$t=386~\mathrm{s}$ the reflected wave reaches the TR. After that time the wave
continues in propagation to higher layers of the solar atmosphere, where we
show two wavefronts at the time $t=410~\mathrm{s}$. The last shot presents, the
wave reaching the X-point at time $t=460~\mathrm{s}$.

\begin{figure*}[t!]
\centering
\includegraphics[width=7.25in]{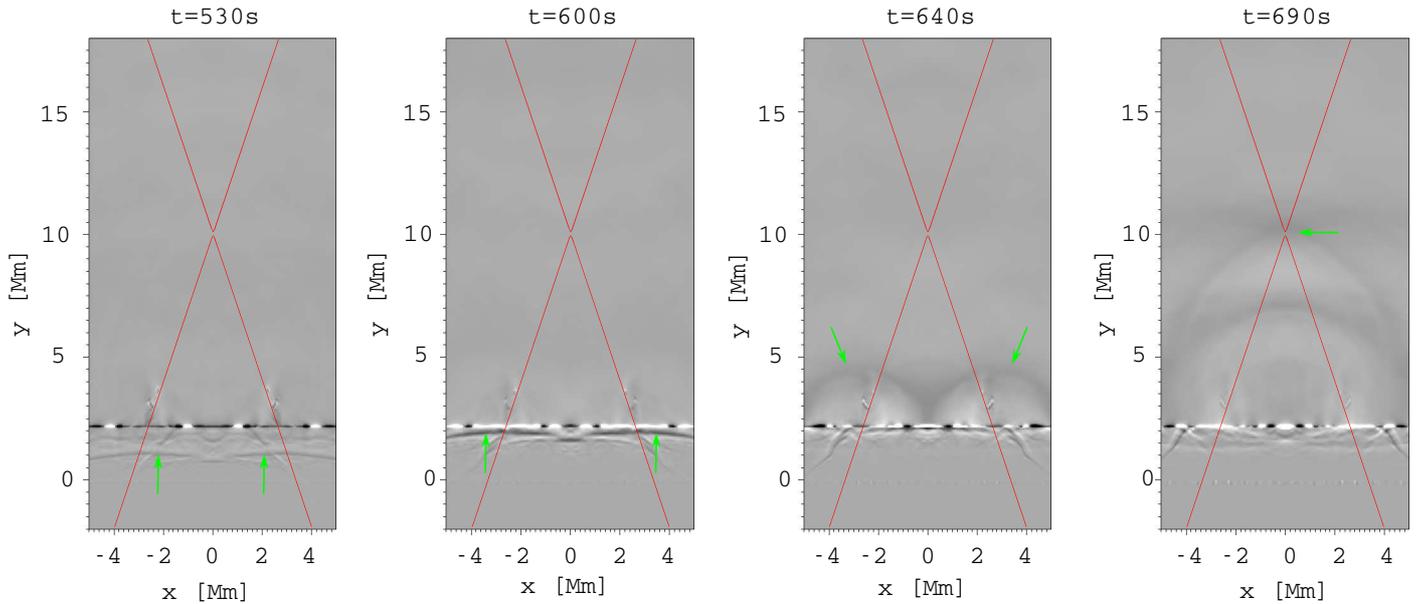}
\caption{Temporal evolution of changes of the mass density
$\Delta\varrho/\varrho_0$, at four different times, $t = 530, 600, 640$ and
$690~\mathrm{s}$, showing the second wave generated by the oscillation of the
photosphere The wave is marked by the lightgreen arrows, the red line
represents the separatrices.} \label{fig4}
\end{figure*}

Fig. {\ref{fig4}} presents again the time evolution of relative change of mass
density for different times $t=530, 600, 640~\mathrm{s}$ and
$t=690~\mathrm{s}$. Similarly as in previous Figure {\ref{fig3}}, we can see
here two waves triggered by oscillating photosphere ($t=530~\mathrm{s}$),
whereas later in time $t=600~\mathrm{s}$ the waves reach the TR. At time
$t=640~\mathrm{s}$ both wavefronts reach each other and continue in propagation
higher to the solar atmosphere. At time $t=690~\mathrm{s}$ they cross the
X-point at height $10~\mathrm{Mm}$. The most important aspect of this set
of pictures is that the initial pressure pulse triggers oscillations of the
photosphere and thus generating waves propagating upwards, where they can
modulate quasi-periodically the process of the flare magnetic reconnection.

\begin{figure}[t!]
\hspace{-0.4cm}
\includegraphics[width=4.0in]{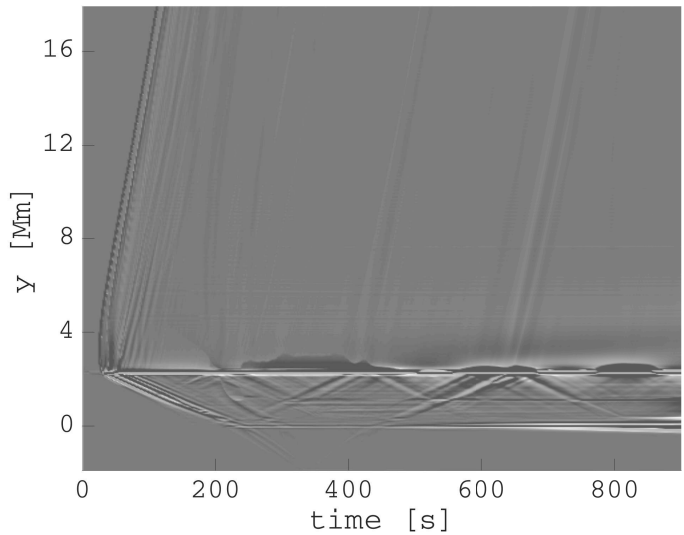}
\caption{The time-space plot showing the evolution of relative mass density change $\Delta\varrho/\varrho_0$ along the $y$-axis of symmetry of the magnetic structure; $x=0~\mathrm{Mm}$.}
\label{fig5}
\end{figure}

\begin{figure}[t!]
\hspace{-0.4cm}
\includegraphics[width=4.0in]{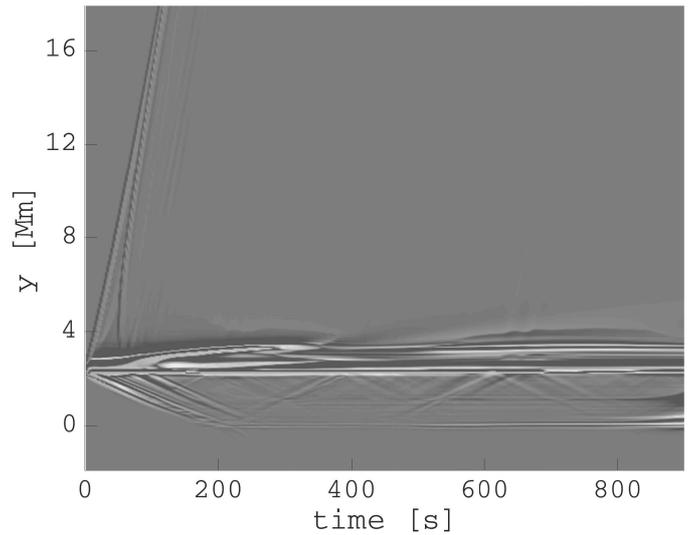}
\caption{The time-space plot showing the evolution of relative mass density change $\Delta\varrho/\varrho_0$ along the $y$-axis passing the perturbation point at $x=2.58~\mathrm{Mm}$.}
\label{fig6}
\end{figure}
In Figs. {\ref{fig5}} and {\ref{fig6}} we show the time-space plot of an
evolution of the relative change of mass density $\Delta\varrho/\varrho_0$ along the
$y$-axis for two values of $x$. In Fig. {\ref{fig5}} it is along the axis of
the symmetry of the magnetic structure, i.e. for $x=0~\mathrm{Mm}$, whereas
Fig. {\ref{fig6}} shows the plot through the perturbation point at
$x=2.58~\mathrm{Mm}$. In both figures, at the beginning, we can see two
shocks (from two spatially separated pulses) propagating upwards and downwards
to higher and deeper layers of the solar atmosphere, respectively.

Arrival of the shocks is delayed because of the distance from the point of
perturbation. The first shock propagates from the nearby perturbation point and
the second one from the more distant perturbation point. The vertical velocity
and the Mach number of the shock propagating upwards are $v_\mathrm{up} \approx
0.18~\mathrm{Mm \cdot s^{-1}}$ and $M \approx$ 1.1 and those of the shock
propagating downwards are $v_\mathrm{down} \approx 0.011~\mathrm{Mm \cdot
s^{-1}}$ and $M \approx$ 1.5. The shock propagating to deeper layers of the
solar atmosphere is partially reflected at time around $t \approx
200~\mathrm{s}$ and changes to waves. Some waves also propagate below the
photosphere to locations, where they are later reflected. The
wave above the photosphere reflects several times between TR and photosphere,
and partially penetrates through TR and propagates upwards into the solar
corona. Figure {\ref{fig6}} is added for comparison. Here, the first shock
starts without any delay, because this plot presents an evolution along the
line which comes through the perturbation point.

\begin{figure}[t!]
\hspace{-0.4cm}
\includegraphics[width=4.0in]{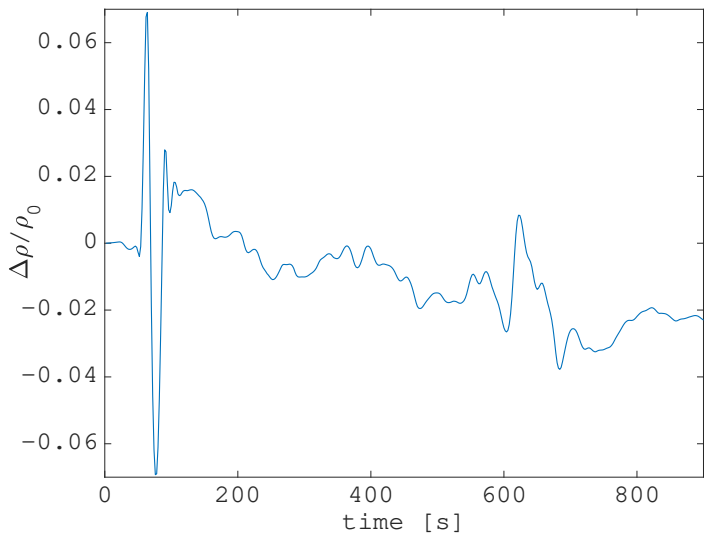}
\caption{The time evolution of relative mass density change $\Delta\varrho/\varrho_0$ in the detection point $x=0~\mathrm{Mm}$ and $y=1.8~\mathrm{Mm}$.}
\label{fig7}
\end{figure}

In Figure {\ref{fig7}} we show the time evolution of relative change of the
mass density in the detection point located at the axis of the symmetry of the
magnetic structure at $y=1.8~\mathrm{Mm}$. (For the position of detection
points, see Figure {\ref{fig1}}.) This figure presents an another view on
shocks (two first peaks) and waves below TR.

\begin{figure}[t!]
\hspace{-0.4cm}
\includegraphics[width=4.0in]{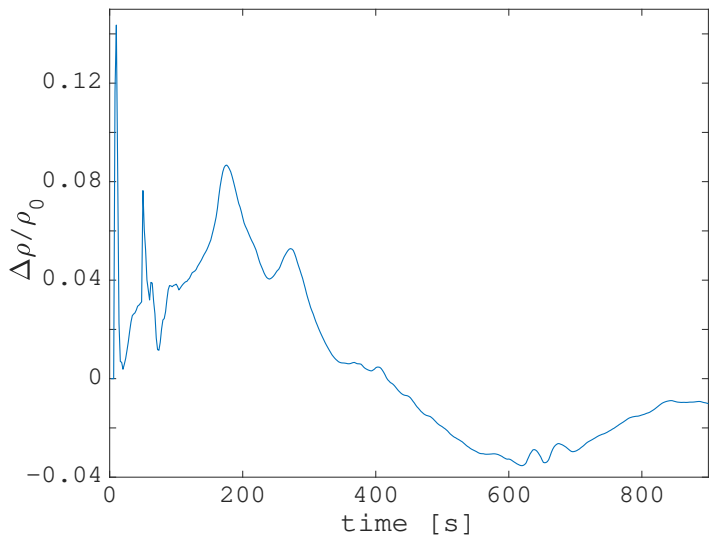}
\caption{The time evolution of relative mass density change $\Delta\varrho/\varrho_0$ at the point $x=2.2~\mathrm{Mm}$ and $y=3.0~\mathrm{Mm}$.}
\label{fig8}
\end{figure}

In Fig. {\ref{fig8}} we present again the time evolution of relative changes of
the mass density $\Delta\varrho/\varrho_0$, but for the point close to the
initial perturbation point, $x=2.2~\mathrm{Mm}$ and $y=3~\mathrm{Mm}$. This
figure shows that during the pulse heating not only shocks are generated, but
also the plasma is evaporated. We can see that after first strong peak at
$t=10~\mathrm{s}$, which corresponds to the shock generated in the nearby
perturbation point, the second peak produced by the shock from the distant
perturbation point appears at $t=50~\mathrm{s}$. This second peak is is
approximately twice ($\Delta=0.1436/0.7637$) lower than the first one. Then at
$t=166~\mathrm{s}$ we observe the peak which is comparable with the previous
one, but it is much broader. This peak shows an arrival of the evaporated
plasma from the nearby perturbation point to the detection one, see also
indication of the plasma evaporation in Fig. {\ref{fig2}}, last two panels for
times $t = 64$ and $106~\mathrm{s}$.

\begin{figure}[t!]
\hspace{-0.4cm}
\includegraphics[width=4.0in]{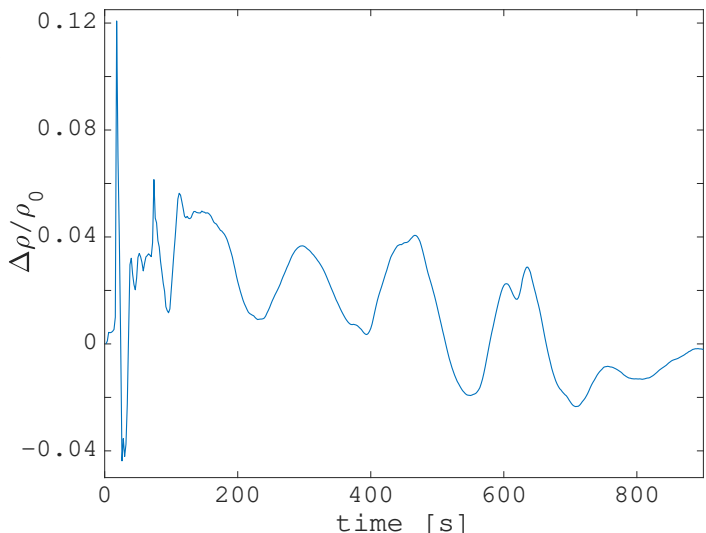}
\caption{The time evolution of relative mass density change $\Delta\varrho/\varrho_0$ at the point $x=4.0~\mathrm{Mm}$ and $y=2.25~\mathrm{Mm}$.}
\label{fig9}
\end{figure}

The last Figure {\ref{fig9}} shows the time evolution of the density
variation at the point $x=4.0~\mathrm{Mm}$ and $y=2.25~\mathrm{Mm}$, which is
located at the same altitude in TR as the perturbation point, but out of it.
This figure shows the wave propagating from the perturbation point through TR.
This wave can correspond to the waves observed sometimes in $H_{\alpha}$ or
UV.


\section{Conclusions}

In this paper we numerically study shocks triggered by pressure pulses launched
in the transition region and consequently generated oscillations of deep
atmospheric layers and waves propagating upwards which can modulate
quasi-periodically the magnetic reconnection in solar flares.
The model is two-dimensional considering gravitationally
stratified solar atmosphere with real temperature distribution according to the
VAL-C model.

The processes shown in this paper represent a new possibility how to explain
some periodicity of solar flare emissions. It is a new alternative to the model
with the resonating loop presented by ~\cite{2006A&A...452..343N}.

A question arises if there are some observational evidence that oscillations of
deep layers of the solar atmosphere produce upwards propagating magnetoacoustic
shock/waves modulating the magnetic reconnection and thus periodically varying
flare emissions.

From our computations it is evident that the oscillations of deep layers are
triggered by a very localized and strong pulse beam heating. The same process
was also proposed for generating of the  seismic
waves~\citep{1998Natur.393..317K,2007ApJ...664..573Z}. Thus, we expect that the
flares associated with the seismic waves could be also connected with the
processes presented in this paper. We plan to analyze this possibility in a
future work.

\begin{acknowledgements}
The authors thank the unknown referee for constructive comments that improved the paper. The support from Grant 16-13277S and M. K. acknowledges also the support from Grants 17-16447S and 19-09489S of the Grant Agency of the Czech Republic is also acknowledged. The authors also express their thanks to Professor Krzystof Murawski for valuable discussions and P. J. thanks of his financial support at UMCS in Lublin, where he worked on this paper during his stay. The FLASH code used in this work was developed by the DOE-supported ASC/Alliances Center for Astrophysical Thermonuclear Flashes at the University of Chicago.
\end{acknowledgements}

\bibliographystyle{aa}
\bibliography{JKosci}

\end{document}